\documentclass[a4paper,11pt]{article}
\usepackage{pos}
\usepackage{xspace}

\usepackage[left]{lineno}

\newcommand{\sib}[1]{\textsc{Sibyll}\,#1\xspace}
\newcommand{\qgsii}{\textsc{QGSJet~II-04}\xspace}
\newcommand{\eposlhc}{\textsc{Epos-LHC}\xspace}
\newcommand{\Xmax}{\ensuremath{X_\text{max}}\xspace}
\newcommand{\DeltaXmax}{\ensuremath{\Delta\Xmax}\xspace}
\newcommand{\EeV}{\,\text{EeV}\xspace}
\newcommand{\eV}{\,\text{eV}\xspace}
\newcommand{\gcm}{\,\text{g}/\text{cm}^{2}}
\newcommand{\avg}[1]{\ensuremath{\langle{#1}\rangle}}

\title{Consequences of a Heavy-Metal Scenario of Ultra-High-Energy Cosmic Rays}

\author*[a]{Jakub Vícha}
\author[a]{Alena Bakalová}
\author[a]{Olena Tkachenko}
\author[a]{Ana Laura Müller}
\author[b,c]{Maximilian Stadelmaier}

\affiliation[a]{FZU - Institute of Physics of the Czech Academy of Sciences, Na Slovance 1999/2, 182 00, Prague 8, Czech Republic}
\affiliation[b]{Università degli Studi di Milano, Dipartimento di Fisica \& INFN, Sezione di Milano, Via Celoria 16, 20133 Milano, Italy}
\affiliation[c]{Karlsruhe Institute of Technology, Institut für Astroteilchenphysik, Hermann-von-Helmholtz-Platz 1
76344 Eggenstein-Leopoldshafen, Germany}

\emailAdd{vicha@fzu.cz}

\abstract{We assume an extreme scenario, in which the arriving cosmic rays are composed of only iron nuclei at energies above $10^{19.6}\,\text{eV}\simeq40\,\text{EeV}$, while allowing a freedom in the scale of the
depth of shower maximum ($X_{\rm{max}}$) and preserving the elongation rate and fluctuations of $X_{\rm{max}}$ predicted by models of hadronic interactions.
We derive the shift of the $X_{\rm{max}}$ scale for \qgsii and \sib{2.3d} models using the public data from the Pierre Auger Observatory.
We then propose a new mass-composition model for the energy evolution of four primary species at the ultra-high energies by fitting the publicly-available $X_{\rm{max}}$ distributions.
We discuss the consequences of this Heavy-metal scenario on the energy spectrum of individual primary species, hadronic interaction studies, and the effect of the Galactic magnetic field on the arrival directions.}

\FullConference{%
$7^\text{th}$ International Symposium on Ultra High Energy Cosmic Rays (UHECR2024)\\
  17. -- 21. November 2024\\
  Malargüe, Mendoza, Argentina\\[4mm]
\includegraphics[width=10em]{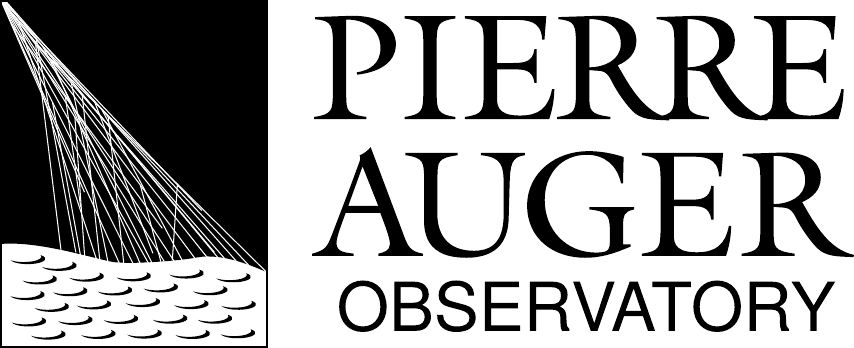}
}

\begin{document}
\maketitle

\section{Introduction}

A consistent interpretation of all measured properties of ultra-high-energy cosmic rays (above ${\sim}10^{18}\eV$) at the Pierre Auger Observatory~\cite{PierreAuger:2015eyc} is difficult to achieve mainly due to the large uncertainties on the mass composition of cosmic rays. 
The current models of hadronic interactions, \eposlhc~\cite{EposLHC}, \qgsii~\cite{Qgsjet}, \sib{2.3d}~\cite{Sibyll}, produce different absolute expectation values for the mass-dependent depth of shower maximum, \Xmax, with the two extremes given by \qgsii and \sib{2.3d}, resulting in the lightest and heaviest mass composition, respectively.
Comparing against expectations from \sib{2.3d}, the measured \avg{\Xmax}~\cite{Auger-LongXmaxPaper,PierreAuger:2023kjt} is approximately at the level of predicted pure nitrogen nuclei at the highest energies, while the measured \Xmax fluctuations are at the level of predictions for pure iron nuclei.
The \Xmax fluctuations predicted by \eposlhc model for iron nuclei are too low due to an overestimation of nuclear defragmentation in this model, which is corrected in the new version of this model~\cite{Pierog:2023ahq} that will soon be available.
Then, the model differences for $\sigma(\Xmax)$ of iron nuclei are within ${\sim}1\gcm$ for all models, and these fluctuations remain robust ($<2\gcm$) even when accounting for modifications of hadronic interactions~\cite{MochiICRC23}.
Recently, the combined measurements of \Xmax and ground signal were found to be well described by models of hadronic interactions, only if the predicted \Xmax scale got deeper for all three models together with an increase of the hadronic part of the ground signal~\cite{PierreAuger:2024neu}.

In this work, we consider a scenario assuming a global shortcoming of all hadronic interaction models to produce the correct \Xmax scale, while predicting correctly the \Xmax fluctuations and elongation rate.
This assumption is motivated by the fact that no consistent interpretation of the mass-dependent observables is possible when comparing to any of the current models of hadronic interactions.
Furthermore, examining the \emph{raw} and uncalibrated results from indirect 
\Xmax estimations, a global shift of $\Xmax$ is apparent~\cite{DeltaMethod,PierreAuger:2024nzw}.
A straight-forward way to solve these issues is to shift the expectation values from hadronic interaction models to a common value, assuming a heavy mass composition.
In the following, we outline the assumptions, methods, and consequences of the heavy-metal scenario for ultra-high-energy cosmic rays.



\section{Heavy-Metal Assumptions}
\label{sec:assumptions}

In the heavy-metal scenario, we make three simple but strong assumptions:
\begin{itemize}
    \item The fluctuations of \Xmax and the elongation rate are correctly described by the \qgsii and \sib{2.3d} models of hadronic interactions,
    \item the absolute \Xmax scale is not correctly described by any of the models and thus there is a freedom to gauge it in a way to make the data be consistently interpretable, and
    \item the cosmic-ray beam consists of pure iron nuclei above energies of $10^{19.6}$\eV (${\sim}40\EeV$).
\end{itemize}

  \begin{figure}
    \includegraphics[width=0.5\textwidth]{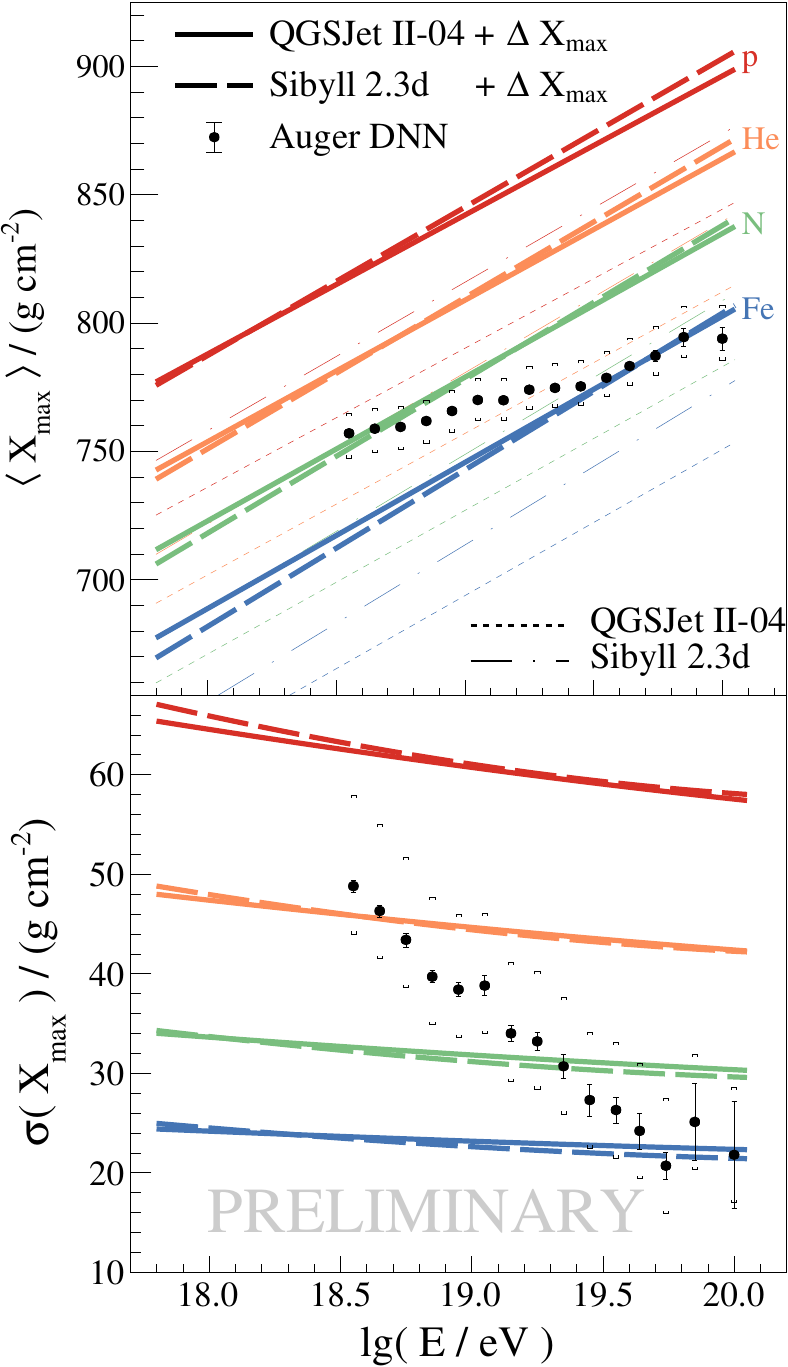}\hspace{0.2cm}
    \includegraphics[width=0.49\textwidth]{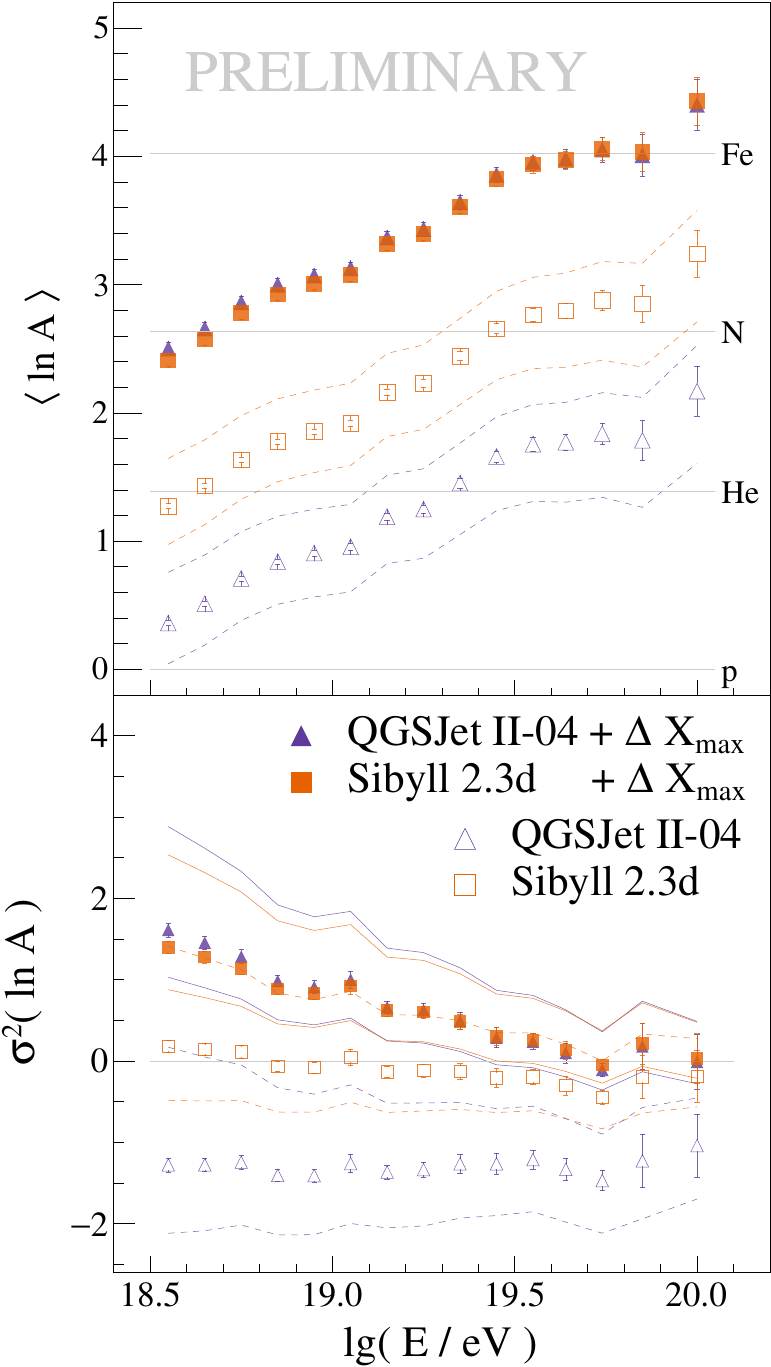}
    \caption{\textbf{Left panel:} Mean and standard deviation of the depth of the shower maximum from \cite{PierreAuger:2024nzw}, systematic uncertainty in brackets, alongside (also shifted by \DeltaXmax) expectations from hadronic interaction models.
    \textbf{Right panel}: Resulting mean and variance of the logarithmic atomic mass of the cosmic-ray primary beam interpreted according to \cite{InterpretationOfXmax} from the \Xmax moments shown in the left panel. Systematic uncertainties are indicated as lines.
    }
    \label{fig:XmaxAndLnAmoments}
  \end{figure}

The measured \Xmax fluctuations reported by the Pierre Auger Collaboration above $10^{19.6}\,\eV$~\cite{PierreAuger:2024nzw} are consistent with the expectations for pure iron nuclei with p($\chi^{2}>0.5$) for both models, see the left panel of Fig.~\ref{fig:XmaxAndLnAmoments}.
Therefore, with these three assumptions at hand, we examine the data of the Pierre Auger Observatory in a holistic way.
Firstly, we determine the modification of the $\Xmax$ scale that is required for each of the models to fit the data in the heavy-metal scenario.
Secondly, we derive the resulting primary fractions as a function of the primary energy, again for both models individually.
Lastly, we discuss the phenomenological aspects of the data in the interpretations of the cosmic-ray energy spectrum for individual primary species, the muon problem in simulated air showers, and backtracking the arrival directions of the most energetic cosmic rays in the Pierre Auger Observatory data; in this context we also discuss the dipolar distribution of cosmic-ray arrival directions above $8\,\EeV$.

\section{Adjustment of \Xmax scale}
\label{sec:adjust}

We use the \Xmax moments derived using Deep Neural Network (DNN) applied to the data recorded by the surface detectors of the Pierre Auger Observatory providing much larger event statistics than the \Xmax moments measured by the fluorescence telescopes~\cite{PierreAuger:2024nzw}.
The measured \Xmax fluctuations are so restrictive that the pure nitrogen nuclei above $10^{19.6}\,\eV$ are in tension with predictions for \qgsii and \sib{2.3d}, see the left panel of Fig.~\ref{fig:XmaxAndLnAmoments}.
We then fit the energy evolution of \avg{\Xmax} above $10^{19.6}$\eV using the prediction of these two models for iron nuclei allowing a shift of the \Xmax scale: $\Xmax \rightarrow \Xmax + \DeltaXmax$.
We obtained $\DeltaXmax=52\pm1^{+11}_{-8}\,\gcm$ and $\DeltaXmax=29\pm1^{+12}_{-7}\,\gcm$ for \qgsii and \sib{2.3d}, respectively.
These values are consistent with the fits of the two-dimensional distributions of \Xmax and ground signal in the energy range $3-10\,\EeV$~\cite{PierreAuger:2024neu}.

In the right panel of Fig.~\ref{fig:XmaxAndLnAmoments}, we plot the interpreted moments of the logarithmic mass number (ln~A), according to \cite{InterpretationOfXmax}, for original model predictions and after correcting the model predictions by applying the \DeltaXmax values.
Note that the original model predictions provided negative variance of ln~A ($\sigma^{2}(\ln \text{A})$) for \qgsii and values consistent with expectations for a pure beam of nuclei in case of \sib{2.3d}.
At the same time, $\avg{\ln \text{A}}$ shows an increasing trend.
After applying a shift of \DeltaXmax, the \Xmax moments can be interpreted in a consistent manner, which is also in agreement with the results of the model-independent estimation of $\sigma^{2}(\ln \text{A})\approx0.7-2.5$ from the correlation between \Xmax and ground signal in the energy range $3-10\,\EeV$~\cite{PierreAuger:2016qzj}.

  \begin{figure}
    \includegraphics[width=0.5\textwidth]{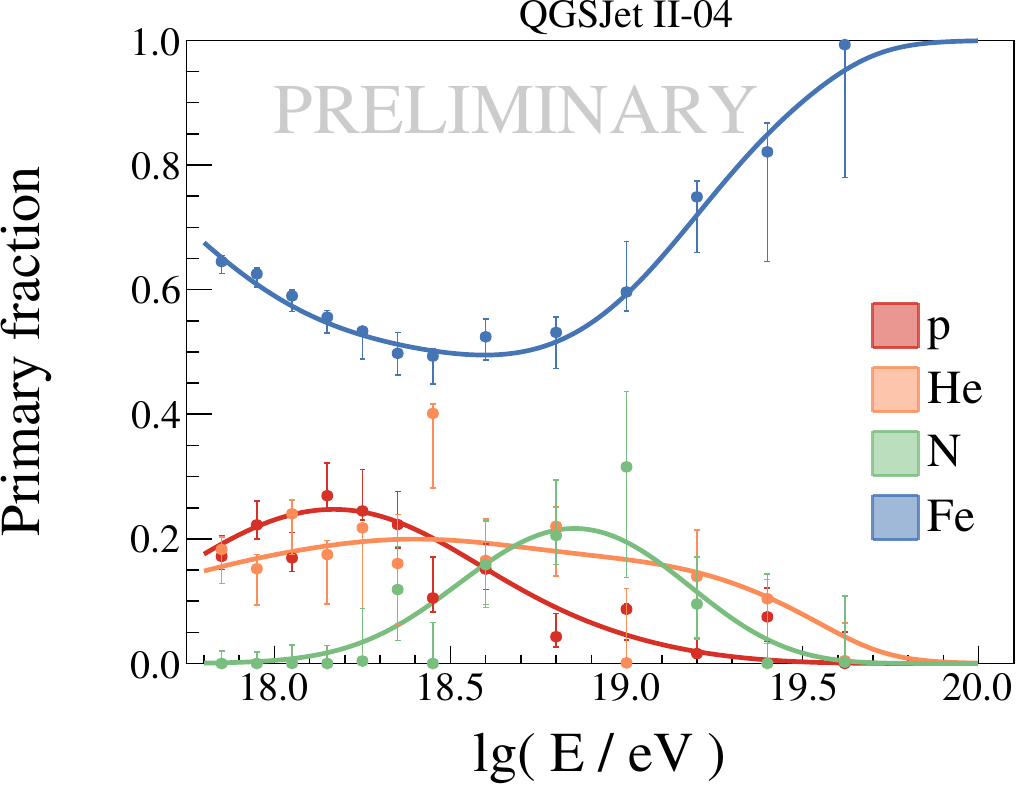}
    \includegraphics[width=0.5\textwidth]{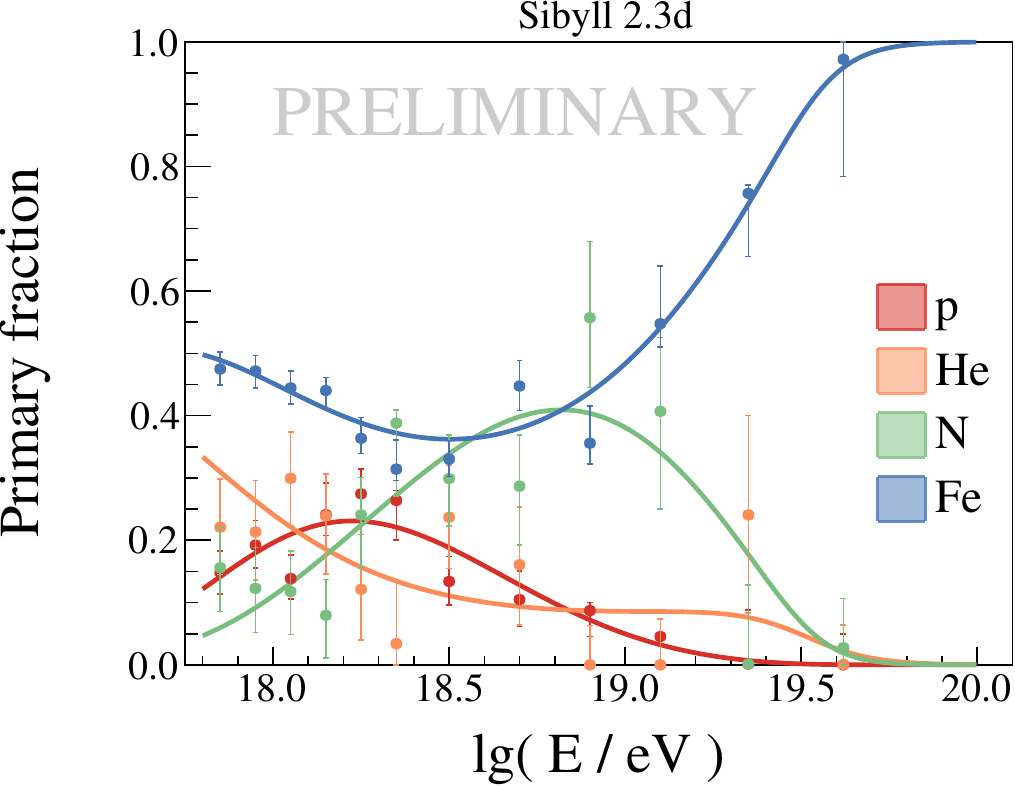}\\ [1cm]
    \includegraphics[width=0.5\textwidth]{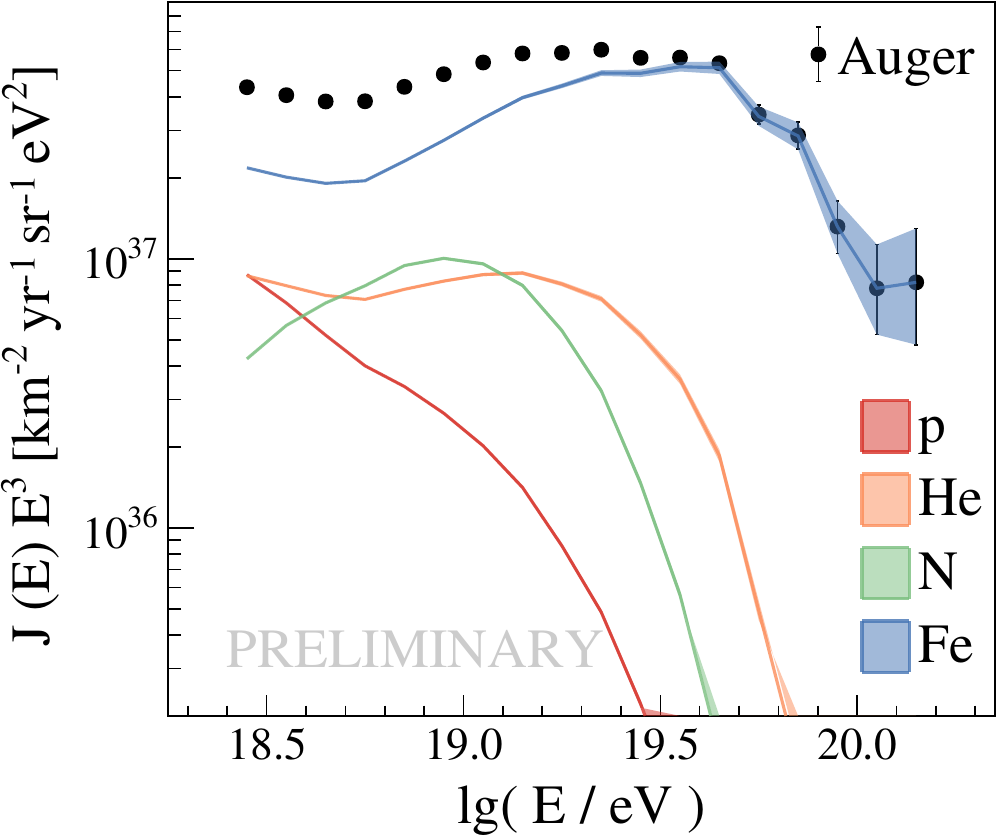}
    \includegraphics[width=0.5\textwidth]{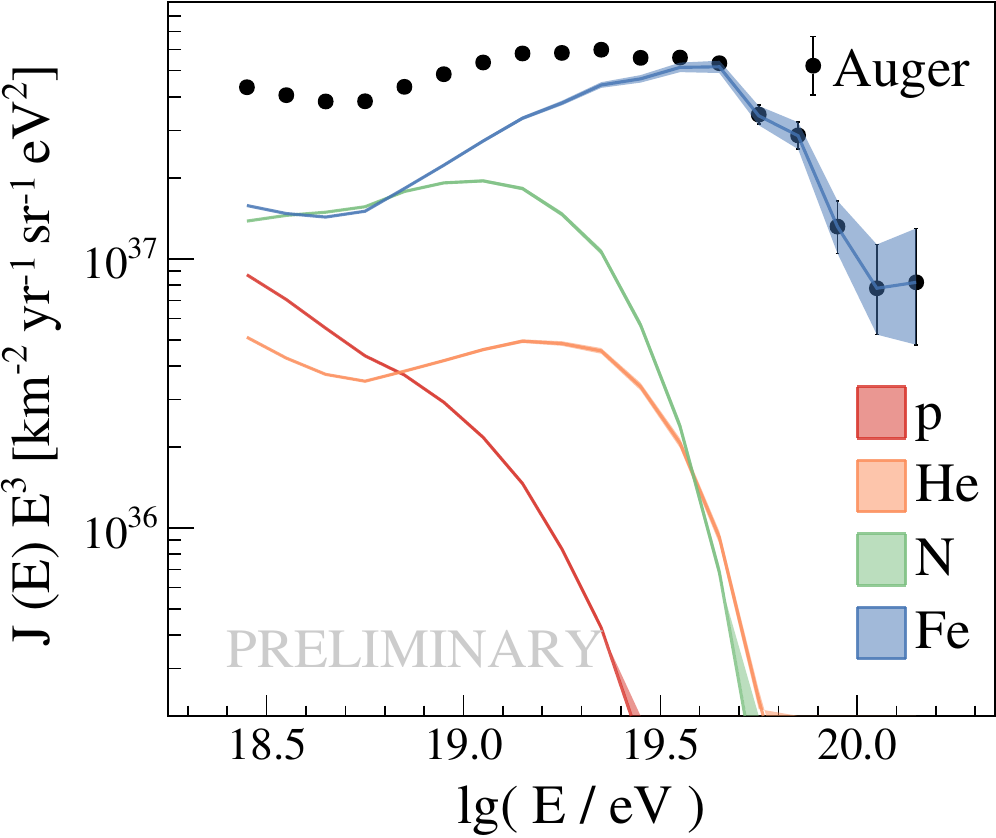}
    \caption{{\bf Top panels:} The energy evolutions of four primary fractions fitted to the \Xmax distributions from \cite{Auger-LongXmaxPaper} together with curves corresponding to their parametrizations for \qgsii (left) and \sib{2.3d} (right). {\bf Bottom panels:} The differential flux measured at the Pierre Auger Observatory \cite{SDEnergySpectrum2020} and contributions of individual primaries derived from the parametrizations shown in the respective top panels.}
    \label{fig:PrimFracVsLogEnergyAndFlux}
  \end{figure}
  
\section{Mass composition}

We consider the \Xmax distributions measured by the Pierre Auger Observatory \cite{Auger-LongXmaxPaper} and apply the shift \DeltaXmax as derived in the previous section for the two models.
Using template distributions for p, He, N, and Fe nuclei obtained from simulations corrected for the detector effects according to \cite{Auger-LongXmaxPaper} we then fit the data to receive relative primary fractions as a function of the primary energy, see the top panels of Fig.~\ref{fig:PrimFracVsLogEnergyAndFlux}.
We use the Gaussian function multiplied by an exponential function (except for iron nuclei, where no exponential is applied) to parametrize these energy evolutions simultaneously.
The resulting sum of parameterizations is normalized to 1 during the log-likelihood minimization.

\section{Energy Spectrum}
In the bottom panels of Fig.~\ref{fig:PrimFracVsLogEnergyAndFlux}, we use the parametrized energy evolutions of the primary fractions from the previous section to derive the flux of individual species from the total energy spectrum measured by the Pierre Auger Observatory~\cite{SDEnergySpectrum2020}. 

Interestingly, the instep feature around 15\EeV can be related to the fading of nitrogen nuclei from the cosmic-ray beam, above which energy the domination of iron nuclei starts to prevail.
The flux suppression at the highest energies is by assumption associated to the iron nuclei.
The suppression of iron and nitrogen nuclei starts approximately at the same rigidity ($E/Z$), which might suggest a common origin of these nuclei.

\section{Hadronic Interactions}
\begin{figure}
    \includegraphics[width=0.49\textwidth]{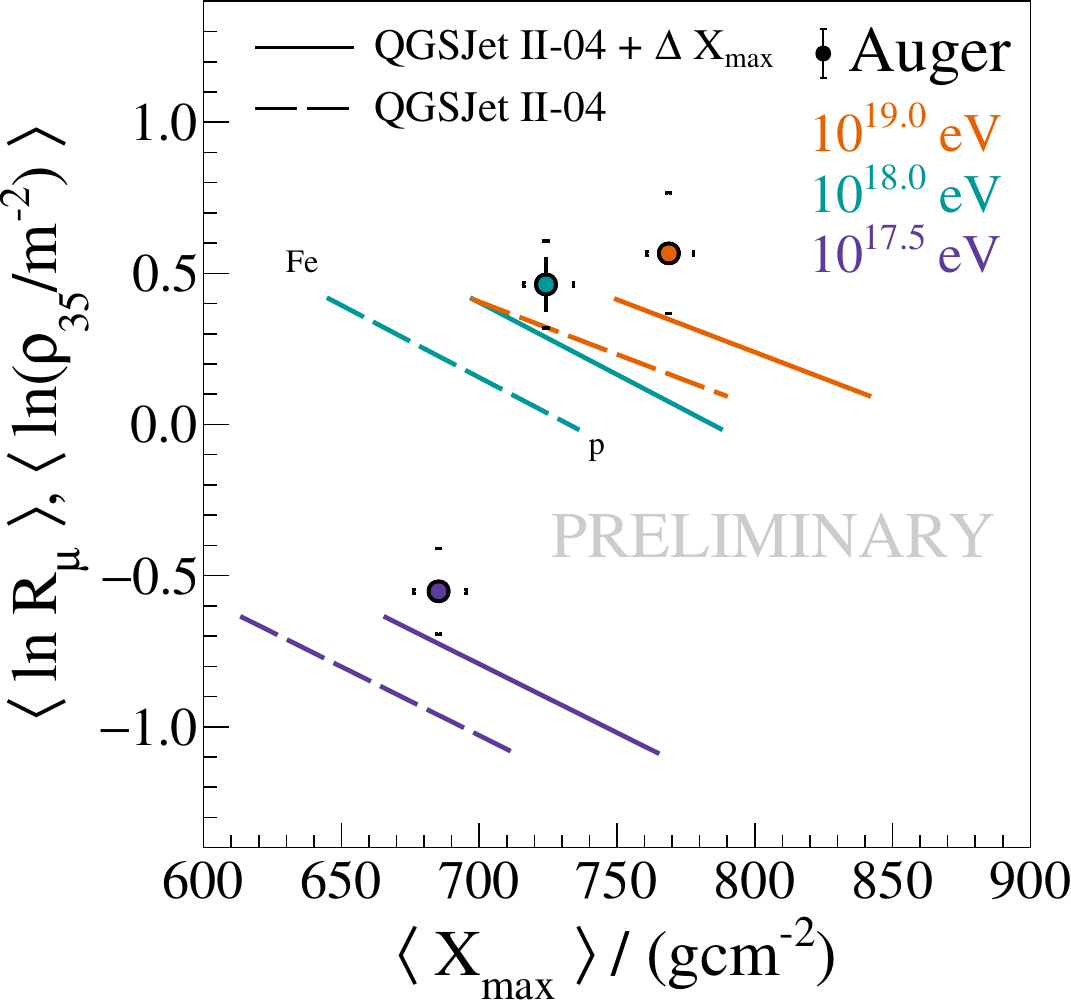}
    \includegraphics[width=0.505\textwidth]{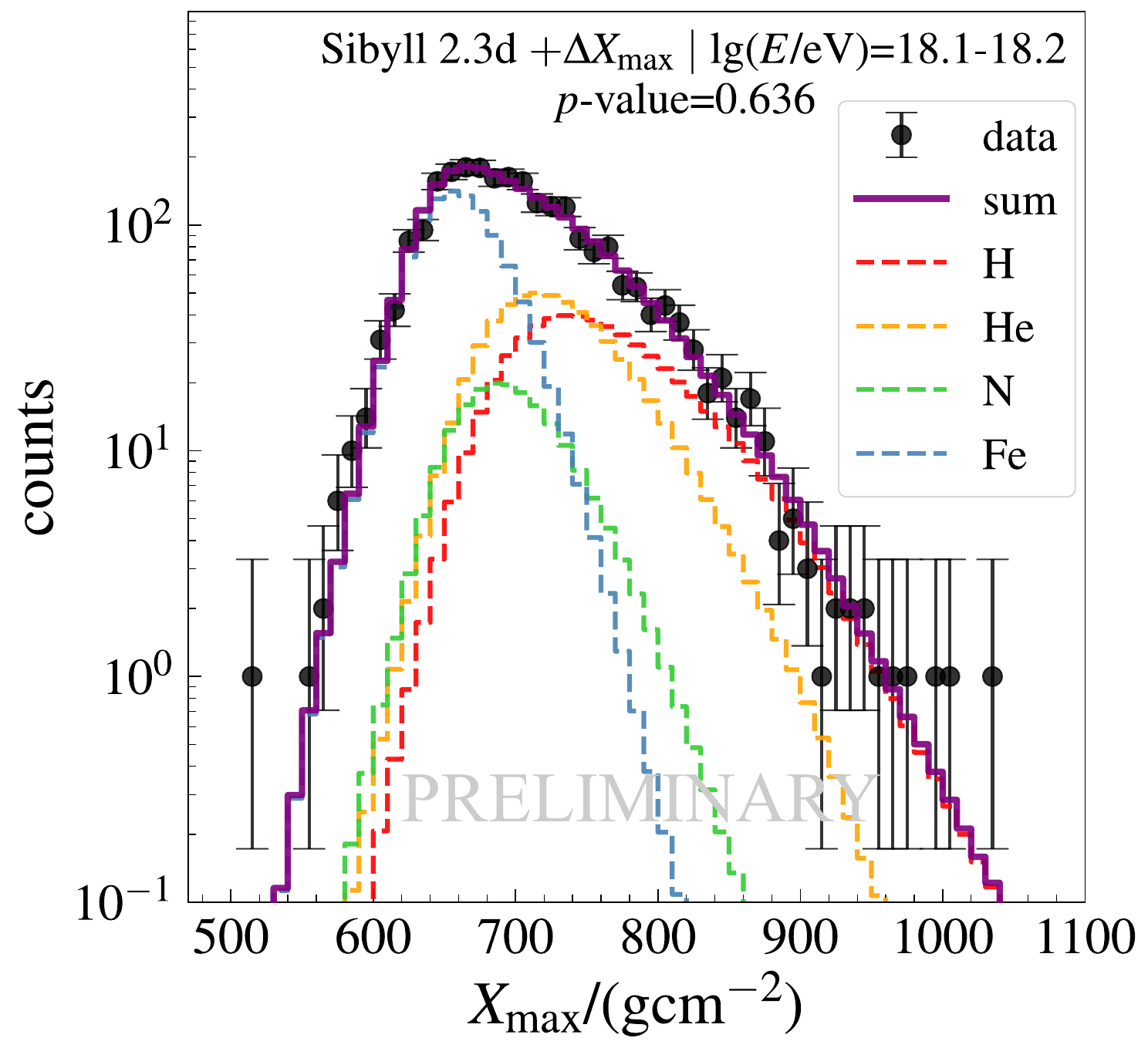}\\ [1cm]
    \includegraphics[width=0.5\textwidth]{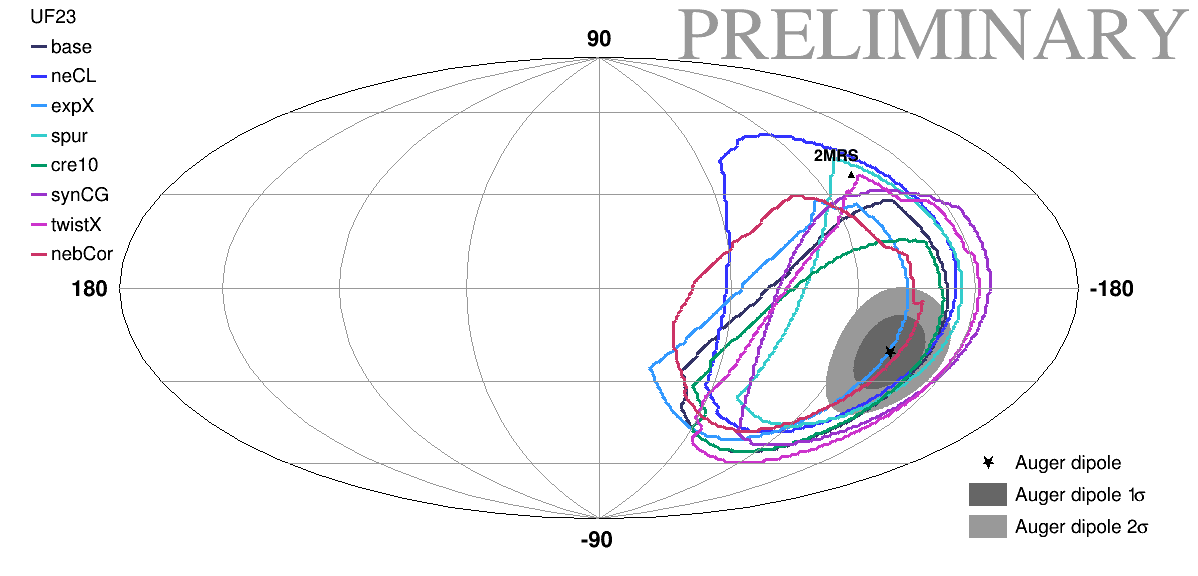}
    \includegraphics[width=0.5\textwidth]{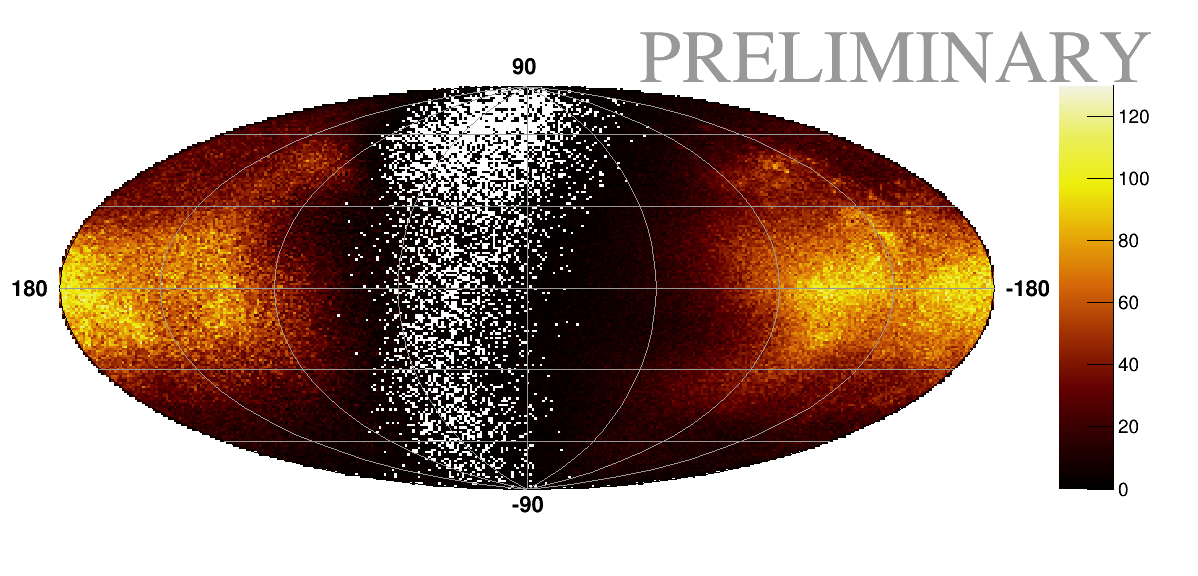}
    \caption{ \textbf{Top-left panel:} The estimations of the muon signal measured directly at the Pierre Auger Observatory for three different energies \cite{MuonFluct2020,AmigaMuons}. We show also the expectations for \qgsii between protons and iron nuclei (dashed lines), and after applying \DeltaXmax (full lines). \textbf{Top-right panel:} An example of the primary-fraction fit to the \Xmax distribution measured by the Pierre Auger Observatory \cite{Auger-LongXmaxPaper} using templates prepared for \sib{2.3d} after applying the detector effects and \DeltaXmax. \textbf{Bottom-left panel:} Identified possible extragalactic directions of a dipole that are consistent within $2\sigma$ with the amplitude and direction of the dipole in the arrival directions measured by the Pierre Auger Observatory \cite{PierreAuger:2017pzq} after propagation in the GMF using UF23 models. \textbf{Bottom-right panel:} Distribution of the directions of 89 cosmic rays above $78\,\EeV$ that were measured by the Pierre Auger Observatory \cite{PierreAuger:2022qcg} after being backtracked to the edge of the Galaxy using the UF23 models of the GMF.}
    \label{fig:HadronicsAndBacktracking}
  \end{figure}

The muon problem for \qgsii identified in direct measurements of the muon signal at the Pierre Auger Observatory~\cite{AmigaMuons,MuonFluct2020} is alleviated from ${\sim}50\%$ to ${\sim}20\%$ when the model predictions of \Xmax are shifted by \DeltaXmax, see the top-left panel of Fig.~\ref{fig:HadronicsAndBacktracking}. This alleviation of the muon problem is consistent with the results from~\cite{PierreAuger:2024neu} with no indication of a non-constant energy evolution of the discrepancy between the model predictions and measured data.

In the top-right panel of Fig.~\ref{fig:HadronicsAndBacktracking}, we show an example of the \Xmax distribution in the energy range $10^{18.1-18.2}\eV$ fitted with the sum of predictions for four primary species generated using \sib{2.3d} that were shifted by \DeltaXmax and corrected for the detector effects according to \cite{Auger-LongXmaxPaper}.
Very good description of the \Xmax distribution has been achieved including the tail of the distribution.
Therefore, there is no indication to modify the elasticity or cross-section of the first interactions to describe especially the tail of measured \Xmax distribution on top of the application of \DeltaXmax.

\section{Backtracking of Arrival Directions}
We backtrack the arrival directions of 89 cosmic rays of energy above $78\,\EeV$ and zenith angle within $60^\circ$ measured by the Pierre Auger Observatory~\cite{PierreAuger:2022qcg} in the Galactic magnetic field (GMF) using the simulation code CRPropa~3~\cite{CRPropa32} as anti-iron nuclei. 
For the GMF, we use the recent UF23 models \cite{UF23} of the coherent component with the turbulent component from the JF12 model \cite{JF12} with the Planck-tuned parameters \cite{JF12Planck}. 
We show the distribution of the backtracked directions at the edge of the Galaxy after re-weighting for the relative exposure of the surface detector of the Pierre Auger Observatory \cite{SOMMERS_SDexposure} in the bottom-right panel of Fig.~\ref{fig:HadronicsAndBacktracking}.
Very similar distribution is obtained for backtracking of an isotropic distribution of arrival directions on the Earth.
The map indicates that in the heavy-metal scenario, we are very limited to measure the cosmic rays at the highest energies from low-longitude directions and most of the cosmic rays are coming from the directions in the Galactic anti-center region.

We are also adopting the method from \cite{BakalovaJCAP23} to restrict the directions of an extragalactic dipole at the $2\sigma$ level that would be consistent in amplitude and direction, after accounting for the effect of the GMF, with the dipole in the arrival directions of cosmic rays above 8\EeV observed by the Pierre Auger Observatory~\cite{PierreAuger:2017pzq}, see the bottom-left panel of Fig.~\ref{fig:HadronicsAndBacktracking}.

\section{Summary}
Given the advent of very precise measurements of \avg{\Xmax} and \Xmax fluctuations above ${\sim}40\,\EeV$ by applying DNN methods to the signals from the surface detectors at the Pierre Auger Observatory, we propose an extreme scenario to describe the ultra-high-energy cosmic rays, yet providing much more consistent interpretation of many measured aspects.
Assuming pure iron nuclei above ${\sim}40\,\EeV$ and freedom in the predicted \Xmax scale we use the model predictions of \qgsii and \sib{2.3d} models to derive the shift of the predicted \Xmax scale providing a consistent interpretation of the two measured \Xmax moments using \avg{\ln A} and $\sigma^{2}(\ln A)$. 
We then derive a mass-composition model of cosmic rays for four primary species that we use to derive individual contributions to the total energy spectrum, effects on the hadronic-interaction studies, and backtracking of arrival directions.

Within the heavy-metal scenario, the instep feature in the cosmic-ray energy spectrum can be explained by an energy threshold for fading of intermediate nuclei from the beam.
The flux suppression at the highest energies is associated to the iron nuclei by assumption, being at the same rigidity as in the case of nitrogen nuclei suggesting their common origin.
The muon problem for \qgsii is alleviated for the cost of the assumed problem in the \Xmax scale, expected to be approximately the same case for other current models of hadronic interactions within our scenario.
There is no indication to change elasticity or cross-section of the first interactions, on top of the application of \DeltaXmax, in the model predictions for \sib{2.3d} to describe better the tail of the measured \Xmax distribution at ${\sim}15\,\EeV$.
Using backtracking of the cosmic rays above $78\,\EeV$ in the GMF, we show that their sources should be mostly in the directions around the Galactic anti-center, while the sources at the low galactic-longitude regions are shadowed by the GMF.
We also show that the dipole anisotropy can be reproduced both in direction and amplitude within the heavy-metal scenario at the $2\sigma$ level.

\acknowledgments
The work was supported by the Czech Academy of Sciences: LQ100102401, Czech Science Foundation: 21-02226M, Ministry of Education, Youth and Sports, Czech Republic: FORTE
CZ.02.01.01/00/22\_008/0004632, and by the PRIME programme of the German Academic Exchange Service (DAAD) with funds from the German Federal Ministry of Education and Research (BMBF).
The authors are very grateful to the colleagues of the Pierre Auger Collaboration for fruitful discussions about this topic.

\bibliographystyle{JHEP_mod}
\bibliography{bibtex.bib}

\end{document}